\documentstyle[12pt]{article}
\begin{document}
\begin{center}
{\large
 NONCOMMUTATIVITY IN MAXWELL-CHERN-SIMONS-MATTER THEORY SIMULATES  PAULI MAGNETIC COUPLING }\\
\vskip 2cm
Subir Ghosh\\
\vskip 1cm
Physics and Applied Mathematics Unit,\\
Indian Statistical Institute,\\
203 B. T. Road, Calcutta 700108, \\
India.
\end{center}
\vskip 3cm
{\bf Abstract:}\\
We study  interactions between like charges in the noncommutative
Maxwell-Chern-Simons electrodynamics
 {\it{minimally}} coupled to  spinors or scalars. We demonstrate that the
 non-relativistic potential profiles, for only spatial noncommutativity,
 are nearly identical to the ones generated by a {\it{non-minimal}} Pauli
 magnetic coupling, originally introduced by
 Stern \cite{js}. Although the Pauli term has crucial roles in the context
 of physically relevant objects such as anyons and like-charge bound states
 (or "Cooper pairs"), its inception \cite{js} (see also \cite{others}) was
 ad-hoc and phenomenological in nature.  On the other hand we  recover similar
 results by extending  the minimal model to the noncommutative plane, which has
 developed in to  an important generalization to ordinary spacetime in recent years.
 No additional input is needed  besides the noncommutativity parameter.

   We prove a novel result that for complex scalar matter sector, the bound
 states (or "Cooper pairs"  can be generated {\it{only}} if the
 Maxwell-Chern-Simons-scalar theory is embedded in noncommutative spacetime.
 This is all the more interesting since the Chern-Simons term does not directly
 contribute a noncommutative correction term in the action.

\newpage
\noindent
{\it Introduction}:\\
In recent times we have learnt to live with the fact that
coexistence of quantum field theory and gravity demands a drastic
change in our notions of geometry, in particular from the
classical spacetime continuum to a quantum fuzzy spacetime
\cite{sz0}. The fuzziness induces a lower bound on the
localization of a spacetime point itself. The need for this length
scale in quantum gravity can be justified from a semiclassical
argument: localization of a particle within the Planck length
requires a large amount of energy that is enough to create a black
hole, which in turn can swallow the particle. This impasse is
avoided by the introduction of a fuzzy spacetime, endowed with an
uncertainty relation of the form
$$\Delta x^i\Delta x^j\geq\frac{1}{2}\mid \theta^{ij}\mid .$$
This phenomenon can be induced by a non-trivial coordinate
commutation relation,
$$[x^i,x^j]=i\theta^{ij},$$
very much in analogy to the standard phase space commutation
relations,
$$[x^i,p_j]=i\hbar\delta^{i}_{j}.$$
The noncommutativity parameter $\theta^{ij}$ plays the role of
Planck's constant $\hbar$.

These heuristic ideas have been strongly supported in string
theory. Ultra-high energy scattering amplitudes suggest a modified
form of Heisenberg phase space uncertainty relation  that directly
leads to a minimum length scale. However, the Non-Commutative (NC)
spacetime scenario has received a great impetus after the seminal
work of Seiberg and Witten \cite {sw}, that relates gauge theories
in NC spacetime to low energy limits of open string theory moving
in an antisymmetric background field \cite{rev}. The inherent
non-locality in NC gauge theories gives rise to a host of
interesting phenomena such as UV/IR mixing \cite{mrs}, loss of
unitarity \cite{gomis} and  violation of Lorentz invariance
\cite{seib} to name a few.

In an alternative approach \cite{das}, one can study effects of
noncommutativity in a {\it local} quantum field theoretic
framework, where one exploits the Seiberg-Witten map \cite{sw}.
This scheme, to be elaborated later, will be followed in our work.

The above discussion is aimed at convincing the reader that, (at
least for distances short enough), NC spacetime is quite natural
and physically motivated {\footnote{It is worth mentioning that a
celebrated example of an NC space appears in the dynamics of
charged particles confined in a plane with a large perpendicular
magnetic field. At the lowest Landau level, where most of the
particles will remain at low energy, the configuration space space
of the particles is effectively noncommutative, with the inverse
of magnetic field identified as $\theta$. The NC manifolds in open
string boundaries emerge in a similar way. }}. In this background,
study of NC extensions of well studied quantum field theories in
ordinary spacetime has gained a lot of importance and the present
work falls in this category, where we will demonstrate that NC
effects alter the behavior of charged matter coupled to
Maxwell-Chern-Simons gauge theory in interesting and non-trivial
ways. In these instances, our approach of analyzing the scattering
and static potential problems will provide further insights on NC
effects in a more familiar (and possibly simpler) setting of
non-relativistic Schrodinger equation formalism.

Let us briefly mention the relevance of Chern-Simons theory in
ordinary spacetime. Arbitrary or "anyonic" statistics \cite{wil}
is a very general consequence of $2+1$-dimensional dynamics since
a non-trivial linking number  can be attributed to the particles
moving in the plane. This exotic statistics is due to  the fact
that the little group of the Poincare group acting on massive
representations is abelian.

This phenomenon was realized in the context of Chern-Simons (CS)
gauge theory \cite{wil} where CS term induces the anyonic behavior
by  attaching a localized magnetic flux to a point charge.
However, it was pointed out \cite{kogan} that it would be more
realistic (for dynamical aspects) to consider the pure CS gauge
theory as a descendent of the topologically massive planar
electrodynamics \cite{dj} (or Maxwell -Chern-Simons (MCS) theory),
in the long wavelength limit. For example, the mass term in
fermionic theory in 2+1-dimensions is parity violating and
sometimes it becomes convenient to introduce the CS term to
represent fermions in terms of bosons \cite{bos}, at least in a
non-relativistic approximation.

Later  a new approach to generate anyons was established by Stern
\cite{js} (see also \cite{others}) that does not require the CS
term, but introduces a generalized connection $\sim
A_{\mu}^{(\kappa)}=A_\mu
+\kappa\epsilon_{\mu\nu\lambda}F^{\nu\lambda}$ with which the
conserved $U(1)$ current is coupled $(eJ^{\mu}A^{(\kappa)}_{\mu})$
in a gauge invariant way. This new parameter $\kappa$ is the
non-minimal Pauli magnetic coupling which is essentially
{\it{phenomenological}} in nature. {\footnote{Note that  the
$2+1$-dimensional (Dirac) $\gamma $-matrix algebra allows one to
include a magnetic coupling without introducing spin degrees of
freedom \cite{js}.}}

 In the present work, we provide a physically motivated alternative to the ad-hoc introduction
 of the Pauli term: Extension of the minimal interaction model in
 the Non-Commutative (NC) plane. We explicitly demonstrate that the
 electron-electron{\footnote{"Electron" is the generic name of a charged particle.}}
 potential
 in a non-relativistic Maxwell-Chern-Simons (MCS) gauge theory  of charged particles
 in the NC
  plane, is same as the potential obtained \cite{gw} in an MCS theory with non-minimal
  (Pauli)
  interaction. It is important to note that our model consists of minimal coupling only
  and so no phenomenological parameter is introduced. The role of the Pauli coupling $\kappa$ is
  taken over by $\theta$ - the NC parameter. It is worthwhile to observe that we study the MCS
  theory because, even though classically a pure Maxwell theory can be considered
  in $2+1$-dimensions, radiative corrections will anyway generate a CS term in the quantum
  theory. This approach of studying NC effects in field theory is new and has not been explored so far.

  Besides exhibiting the Pauli magnetic moment effect and anyonic
interaction potential, the NC model we have studied plays an
important role in the formation of "Cooper pairs" of
electron-electron bound states in the plane \cite{gir1,gw}.   In
the context of "Cooper pair" condensation for scalar charges, the
striking result is that the bound states appear {\it{only}} in the
NC extension and not in the ordinary MCS-Scalar theory. This is
all the more intriguing since the CS term does not directly
contribute in the $O(\theta)$ corrected classical action. It
appears in the one-photon exchange M\"{o}ller scattering
amplitude. On the other hand, for spinorial (Dirac) matter, the
possibility of bound state formation is already present in the
MCS-spinor theory and the NC effect introduces a correction to
that. This is very similar to the effect induced by the Pauli term
\cite{gw}. The vital role played by the noncommutative space in
triggering the bound state formation in scalar-MCS theory is a new
result.

We follow the method used in \cite{gw,gir1} where one starts from
the (one photon)  matrix element of M\"{o}ller scattering between
relativistic electrons and subsequently enforces the
non-relativistic limit. Fourier transform of the momentum space
matrix element yields the interaction potential \cite{sak}. We
have restricted our analysis to $O(\theta )$ - the lowest
non-trivial order in noncommutativity.

After introducing the MCS electrodynamics  and its NC extensions
in the context of spinor and scalar matter sectors \cite{gs,jur},
we derive the M\"{o}ller matrix element from which the
inter-particle potential is generated in the low energy limit. We
conclude with a number of exciting areas which need to be looked
at in the present formalism. \vskip .2cm \noindent
{\it Magnetic coupling effects in NC space}:\\
The Dirac particles interacting minimally with MCS theory is
\begin{equation}
{\cal{L}}=\bar\psi(i\gamma^{\mu}D_\mu -m)\psi
-\frac{1}{4}F^{\mu\nu}F_{\mu\nu}
-\frac{s}{4}\epsilon^{\mu\nu\lambda}A_\mu F_{\nu\lambda},
\label{1}
\end{equation}
where
$$D_\mu \psi\equiv (\partial_{\mu}+ieA_{\mu})\psi ,~~ F_{\mu\nu}\equiv \partial_{\mu}A_\nu -\partial_{\nu}A_\mu .$$ $s$ denotes the coupling of the topological term. We now introduce the NC spacetime with its associated $*$-product (or  Moyal-Weyl product):
\begin{equation}
[x^{\rho},x^{\sigma}]_{*}=i\tilde\theta^{\rho\sigma}, \label{nc}
\end{equation}
\begin{equation}
p(x)*q(x)=pq+\frac{i}{2}\tilde\theta^{\rho\sigma}\partial_{\rho}p\partial_{\sigma}q+~O(\tilde\theta^{2}).
\label{mw}
\end{equation}
The NC generalization of (\ref{1}) is,
\begin{equation}
\hat{{\cal L}}=\hat{\bar\psi}(i\gamma^{\mu}\hat D_\mu -m)*\hat\psi
-\frac{1}{4}\hat F^{\mu\nu}*\hat F_{\mu\nu}
-\frac{s}{4}\epsilon^{\mu\nu\lambda}\hat A_\mu *\hat
F_{\nu\lambda}. \label{2}
\end{equation}
The "hatted" variables are the counterparts of the normal
variables living in NC spacetime, with the following
identifications,
$$\hat D_\mu *\hat\psi =(\partial_{\mu}+ie\hat A_{\mu}*)\hat\psi ;~~
\hat F_{\mu\nu}=\partial_{\mu}\hat A_{\nu}-\partial_{\nu}\hat
A_{\mu}-i\hat A_{\mu}*\hat A_{\nu}+i\hat A_{\nu}*\hat A_{\mu}. $$
As is well-known \cite{rev}, the gauge invariance is elevated to
$*$-gauge invariance in the NC plane. It was shown by Seiberg and
Witten \cite{sw} that appearance of noncommutativity is dictated
by the choice of regularization in the quantum theory and quite
naturally the NC version of a theory should be directly related to
the commutative one by a change of variables. Explicit form of
this Seiberg-Witten map \cite{sw} to the lowest non-trivial order
in $\theta $ is,
$$
\hat A_{\mu}=A_{\mu}+\theta^{\sigma\rho}A_{\rho}(\partial_{\sigma}
A_{\mu}-\frac{1}{2}\partial_{\mu} A_{\sigma})~;~ \hat
F_{\mu\nu}=F_{\mu\nu}+\theta^{\rho\sigma}(F_{\mu\rho}F_{\nu\sigma}-A_{\rho}\partial_{\sigma}
F_{\mu\nu}),$$
\begin{equation}
\hat\psi =\psi -\frac{1}{2}\theta^{\mu\nu}A_\mu \partial_{\nu}\psi
. \label{swm}
\end{equation}
We have scaled $e\tilde\theta \equiv \theta $. The map (\ref{swm})
allows us to study NC effects in the framework of commutative
quantum field theory. The important feature of this map is that it
preserves gauge orbits and so $*$-gauge invariance is translated
in to normal gauge invariance.

Thus (\ref{2}) and (\ref{swm}) generates the following theory in
commutative spacetime,
$$
\hat{\cal{L}}=\bar\psi(i\gamma^{\mu}D_\mu -m)\psi
-\frac{1}{4}F^{\mu\nu}F_{\mu\nu}(1+\frac{1}{2}\theta^{\alpha\beta}F_{\alpha\beta})
-\frac{s}{4}\epsilon^{\mu\nu\lambda}A_\mu
F_{\nu\lambda}-\frac{1}{2\alpha}(\partial^{\mu}A_\mu )^2 $$
\begin{equation}
-\frac{1}{4}\theta^{\alpha\beta}F_{\alpha\beta}\bar\psi(i\gamma^{\mu}D_\mu
-m)\psi
-\frac{i}{2}\theta^{\alpha\beta}F_{\mu\alpha}\bar\psi\gamma^{\mu}D_\beta
\psi , \label{3}
\end{equation}
where total derivative terms have been dropped and simplifications
in the $\theta $-term due to the dimensionality being $2+1$  is
taken in to account. Also a gauge fixing $\alpha$-dependent term
is put in.

In complete analogy, from the bosonic model,
\begin{equation}
{\cal{L}}=(D^\mu \phi )^\dagger D_\mu \phi -m^2\phi ^\dagger \phi
-\frac{1}{4}F^{\mu\nu}F_{\mu\nu}
-\frac{s}{4}\epsilon^{\mu\nu\lambda}A_\mu F_{\nu\lambda},
\label{7a}
\end{equation}
one obtains the $O(\theta)$ NC lagrangian,
$$
\hat{\cal{L}}=(D^\mu \phi )^\dagger D_\mu \phi -m^2\phi ^\dagger
\phi
-\frac{1}{4}F^{\mu\nu}F_{\mu\nu}(1+\frac{1}{2}\theta^{\alpha\beta}F_{\alpha\beta})
-\frac{s}{4}\epsilon^{\mu\nu\lambda}A_\mu
F_{\nu\lambda}-\frac{1}{2\alpha}(\partial^{\mu}A_\mu )^2 $$
\begin{equation}
 -\frac{1}{4}\theta^{\alpha\beta}F_{\alpha\beta}[D^\mu \phi )^\dagger D_\mu \phi -m^2\phi ^\dagger \phi ]
+\frac{1}{2}\theta^{\alpha\beta}[F_{\alpha\mu}(D_\beta
\phi)^\dagger D^\mu \phi+(D^\mu \phi)^\dagger D_\beta \phi ].
\label{18}
\end{equation}

Thus in (\ref{3}) and (\ref{18}) we have developed the models for
spinors and scalars respectively, where the NC effects appear as
interaction terms. Our aim is to extract the inter-particle
potential from the non-relativistic limit of the M\"{o}ller
scattering between two fermions at the tree level, considering
single photon scattering only. A generic feature of NC extension
of a field theory is that the free (quadratic) part is not
modified and so one is allowed to use the propagators and free
field solutions of the commutative theory. The topologically
massive photon propagator in momentum space is
$$
<A^{\mu}A^{\nu}>(k)=Ag^{\mu\nu}+Bk^{\mu}k^{\nu}+iC\epsilon^{\mu\nu\lambda}k_\lambda
$$
\begin{equation}
A=\frac{2}{-k^2 +s^2}~;~~B=-\frac{\alpha
A}{k^2}(-\frac{s^2}{k^2}+1-\frac{1}{\alpha})~;~~C=\frac{-sA}{k^2}.
\label{5}
\end{equation}
We first concentrate on the spinor case. In the fermion content,
the $\gamma $-matrices satisfy the $so(2,1)$ algebra
$[\gamma^{\mu},\gamma^{\nu}]=2i\epsilon^{\mu\nu\lambda}\gamma_{\lambda}$.
They represent a $2+1$-dimensional representation of the Dirac
matrices, {\it{i.e.}} the Pauli matrices: $\gamma^{\mu}\equiv
(\sigma_{z}, -i\sigma_{x}, i\sigma_{y})$. The free spinor
solutions are given by,
\begin{equation}
u(p)=\frac{1}{\sqrt{2m(E+m)}} \left(
\begin{array}{c}
E+m \\
-ip_x-p_y
\end{array}
\right )~;~~\bar u(p)=\frac{1}{\sqrt{2m(E+m)}} (E+m
~~~~-ip_x+p_y). \label{4}
\end{equation}
It is convenient to introduce the Gordon identity (in
2+1-dimensions)
\begin{equation}
j^{\mu}(p',p)\equiv \bar
u(p')\gamma^{\mu}u(p)=\frac{2m}{4m^2-k^2}[\bar
u(p')u(p)]((2p-k)^{\mu}+\frac{i}{m}\epsilon^{\mu\nu\lambda}k_\nu
p_{\lambda}), \label{6}
\end{equation}
where $k^\mu \equiv (p'-p)^\mu $. Since we are interested in the
non-relativistic limit, interaction terms with smaller number of
derivatives will dominate. This leads us to a truncated form of
the interaction part:
\begin{equation}
\hat{\cal{L}}_{Int}=-A^\mu\bar\psi\gamma_{\mu}\psi
+\frac{m}{4}\theta^{\alpha\beta}F_{\alpha\beta}\bar\psi \psi .
\label{7}
\end{equation}
The first term is the normal $U(1)$ gauge interaction term whereas
the $\theta $-term is of a Yukawa interaction type where the
massive photon interacts with the fermion mass term.

Already it is apparent that non-locality, in the form of magnetic
moment of the otherwise spinless fermion, will play an essential
role since the $\theta $-contribution of the interaction depends
on the factor $\theta^{\mu\nu}k_\nu $. This is clearly reminiscent
of the dipole nature of the NC Maxwell theory \cite{jab} where the
spatial extent of the dipole is $\sim\theta^{\mu\nu}k_\nu $. The
connection with the phenomenological models \cite{js,others} with
Pauli interaction
$F^{\mu\nu}\bar\psi[\gamma_{\mu},\gamma_{\nu}]\psi $ is also
obvious.

The matrix element of the M\"{o}ller scattering has two parts: the
Coulomb term $(M_I)$ and the $\theta $-term $(M_{II})$, which are
given below,
\begin{equation}
-iM_I=(ie)^2j^\mu (p_1',p_1)j^\nu (p_2',p_2)<A_\mu A_\nu >(k),
\label{8}
\end{equation}
\begin{equation}
-iM_{II}=(ie)\frac{m}{2}\theta^{\alpha\mu}k_\alpha j
(p_1',p_1)j^\nu (p_2',p_2)<A_\mu A_\nu >(k), \label{9}
\end{equation}
where $j (p',p)\equiv \bar u(p')u(p)$. In the relativistic theory
there will be a contribution of the exchange term which is
obtained by interchanging the final state labels and keeping in
mind the particle statistics. However, this is not required for
our purpose since we will study the Schrodinger potential problem
where taking anti-symmetric wave functions will take care of the
effect of the exchange term \cite{sak}. Equivalently, one can
think of the particles as distinguishable. We introduce the center
of mass frame and revert to a non-relativistic notation,
$$k^\mu \equiv (0,\vec{k})~;~~p_1^\mu \equiv (E,\vec{p})~;~~p_2^\mu \equiv (E,-\vec{p})~;~~\theta_{\mu}=\frac{1}{2}\epsilon_{\mu\nu\lambda}\theta^{\nu\lambda}\equiv (\theta_{0},\vec{\theta }).$$
A straightforward computation of the matrix elements yields,
$$M_I=\frac{e^2}{8m^4(E+m)^2}(1+\frac{\vec{k}^2}{4m^2})^{-2}
\frac{1}{ {\vec
{k}^2}+s^2}[2m(E+m)-(\vec{k}.\vec{p})-i(\vec{p}\times
\vec{k})]^2[2iE(\vec{k}\times
\vec{p})\{4(\frac{1}{m}+\frac{s}{\vec{k}^2})-\frac{s}{m^2}\}$$
\begin{equation}
+(p_1.p_2)(4-\frac{\vec{k}^2}{m^2}-\frac{4s}{m})-(\vec{p}.\vec{k})\{4-\frac{s}{m}-\frac{(\vec{p}.\vec{k})}{m}(\frac{1}{m}+\frac{4s}{\vec{k}^2})\}+\vec{k}^2],
\label{10}
\end{equation}
$$M_{II}=-\frac{ie}{2}\frac{1}{\vec{k}^2+s^2}(1+\frac{\vec{k}^2}{4m^2})^{-2}[-(2+\frac{s}{m})(E(\vec{k}\times \vec{\theta})+\theta_{0}(\vec{k}\times \vec{p}))$$
\begin{equation}
-i(\frac{1}{m}+\frac{2s}{\vec{k}^2})\{-(\vec{k}.\vec{p})(\vec{k}.\vec{\theta
}) )+(\theta_0E+(\vec{\theta } .\vec{p}))\vec{k}^2\}]. \label{11}
\end{equation}
$\alpha $-dependent terms will not occur since conserved currents
are involved. The non-relativistic limit simplifies the
expressions considerably and we find,
\begin{equation}
M_I=\frac{e^2}{\vec{k}^2+s^2}[(1-\frac{s}{m}+\frac{2is}{m}\frac{(\vec{k}\times
\vec{p})}{\vec{k}^2}], \label{12}
\end{equation}
\begin{equation}
M_{II}=-\frac{e}{2(\vec{k}^2+s^2)}[\theta_{0}\{\vec{k}^2+2sm-2i(\vec{k}\times
\vec{p})\}-2s{\frac{(\vec{\theta}.\vec{k})(\vec{p}.\vec{k})}{\vec{k}^2}}+2s\vec{\theta
} .\vec{p}-im(2+\frac{s}{m})(\vec{k}\times \vec{\theta})].
\label{13}
\end{equation}
Defining the Fourier transform as
$$V(r)=\frac{1}{(2\pi)^2}\int d^2k~e^{i\vec{k}.\vec{r}}M(k),$$
we immediately obtain the cherished form of the electron-electron
potential,
\begin{equation}
V_I=e^2[\frac{1}{2\pi }(1-\frac{s}{m})K_0(sr)-\frac{1}{\pi
ms}\frac{L}{r^2}(1-srK_1(sr))], \label{14}
\end{equation}
\begin{equation}
V_{II}=-\frac{e}{4\pi}[\theta_{0}\{2smK_0+\frac{2sL}{r}K_1\}+\theta
\{2spK_0
+(\frac{ms}{p}(2+\frac{s}{m})\frac{L}{r}+\frac{4}{p}\frac{L^2}{r^3}-\frac{2p}{r})K_1-\frac{2}{sp}\frac{L^2}{r^4}\}].
\label{15}
\end{equation}
In the above expression, we have chosen $\vec{\theta }=\theta \hat
p$ where $\hat p\equiv \frac{\vec{p}}{p}$ is the unit vector along
$\vec p$. This simply means that we have taken the center of mass
frame in such a way that $\vec{p}$ coincides with the given
constant direction $\vec{\theta}$. The expression of $V_{II}$ in
(\ref{15}) is one of the main results of the present work.

$V_I$ is the potential in commutative spacetime reported before
\cite{gir1} and $V_{II}$ constitute the $O(\theta)$ correction.
For reasons of unitarity \cite{gomis}, in the study of NC quantum
field theory, one generally restricts the noncommutativity to
affect only space coordinates, keeping time as a commutative
parameter. Then one immediately notices that for only spatial
noncommutativity, ({\it{i.e.}} $\vec{\theta } =0,~\theta_{0}\neq
0$ {\footnote{In NC spacetime, $\theta_{0}$ will be determined by
stringy effects. On the other hand, one can think of $\theta_{0}$
as arising from a lowest Landau level scenario, in which case it
will be controlled by the inverse of magnetic field.}}),
$V_{II}(\theta _0)$ does not introduce any structural change
(regarding $r$-dependence) in the potential and one finds the full
potential for the fermions  to be,
\begin{equation}
V(\theta _0)\mid_{spinor}=\frac{e^2}{2\pi
}[1-\frac{s}{m}-\frac{\theta_{0}sm}{e}]K_0(sr)-\frac{e^2}{\pi
ms}\frac{L}{r^2}[1-(1-\frac{\theta_{0}sm}{2e})srK_1(sr)].
\label{16}
\end{equation}

The computations for the bosonic case is simpler. In the low
energy limit, the leading interaction terms are
\begin{equation}
{\cal{L}}_{Int}=(D^\mu \phi )^\dagger D_\mu \phi
+\frac{m^2}{4}\theta^{\alpha\beta}F_{\alpha\beta}\phi ^\dagger
\phi . \label{b1}
\end{equation}
In the non-relativistic limit, we get the $\theta $-contribution
to be,
\begin{equation}
M_{II}=-\frac{2ie}{\vec{k}^2+s^2}[\theta_{0}\{(\vec{k}\times
\vec{q})-ism\}-m(\vec{k}\times
\vec{\theta})-\frac{is}{\vec{k}^2}\{(\vec{k}. \vec{q})(\vec{k}.
\vec{\theta})-\vec{k}^2(\vec{\theta}. \vec{q})\}]. \label{b2}
\end{equation}
This yields the potential for the bosonic case for only spatial
noncommutativity,
\begin{equation}
V(\theta _0)\mid_{scalar}=\frac{e^2}{2\pi
}[1-\frac{\theta_{0}sm}{e}]K_0(sr)-\frac{e^2}{\pi
ms}\frac{L}{r^2}[1-(1-\frac{\theta_{0}sm}{2e})srK_1(sr)].
\label{b3}
\end{equation}
This constitutes the other main result.

Comparing the potential profiles (\ref{16}) and (\ref{b3}), we
immediately spot the crucial difference: in the  expression for
scalar matter in (\ref{b3}), the term $\sim \frac{s}{m}K_0(sr)$ is
missing. This shows that for scalars, the term that is essential
in reversing the Coulomb repulsion for bound state formation, is
generated only in the NC regime. This  rather dramatic outcome of
the NC extension  is a new result.

On the other hand, in case of fermions, the CS term by itself is
able to reverse the normal (logarithmic) Coulomb repulsion between
electrons making it conducive for the formation of
electron-electron bound states. The Pauli non-minimal coupling
generates an additional contribution in the potential that is
similar to the CS contribution. A similar situation prevails in
the present case where NC effects induce an additional term in the
potential that is  similar to the CS contribution.

As mentioned before, exactly similar forms of inter-electron
potentials have been reported before \cite{gw} in the context
{\it{non-minimal}} Pauli coupling proposed in  \cite{js}. The
parameter $\theta_0$, ($\frac{m\theta_{0}}{2e}$ to be precise), is
to be identified with the non-minimal coupling $\kappa$ \cite{gw},
where $\kappa$ is defined in terms of the covariant derivative
$D_\mu\psi\equiv (\partial_{\mu} +ieA_\mu
+i\frac{e\kappa}{2}\epsilon_{\mu\nu\lambda}F^{\nu\lambda})$. There
is a difference  between the potentials in \cite{gw} and
$V({\theta _0})$ in (\ref{16}) and (\ref{b3}), the latter two
receiving a $\theta$-contribution in the angular momentum ($L$)
term as well. An intriguing point is that, although there appears
no explicit contribution of the CS term in the $\theta
$-correction upon exploiting the Seiberg-Witten map \cite{gs}, the
$O(\theta _0)$ correction terms in the potential (\ref{16}) and
(\ref{b3}) are dependent on $s$ and the correction term in $K_0$
will vanish if the CS term is absent. Actually the CS term
converts the photon to a massive one which plies between the
charges. This establishes the fact that the desired results are
obtainable in the NC extension of the MCS-charge model, instead of
incorporating the Pauli term in an ad-hoc way.

For $s$ being small compared to $m$, one can take $K_0(sr)\sim
-ln(sr);~K_1(sr)\sim \frac{1}{sr},$ so that $V({\theta _0})$
reduces to
\begin{equation}
V(\theta_0)\mid_{spinor}\sim \frac{e^2}{2\pi
}[\frac{s}{m}+\frac{\theta_{0}sm}{e}-1]ln(sr)-\frac{e\theta_{0}}{2\pi}\frac{L}{r^2}.
\label{17}
\end{equation}

Assuming $e\theta_{0}$ to be small we neglect the last term and
subsequently can read off an approximate expression for the
$S$-wave binding energy from a semi-classical analysis performed
in \cite{gw}. The result is,
\begin{equation}
E_{n,0}\mid_{spinor}\sim \frac{e^2s}{\pi}\nu ~
ln[\frac{2\pi}{e}(n+\frac{1}{2})\sqrt{\frac{s}{m\nu}}],
\label{17a}
\end{equation}
where
$\nu=\frac{\theta_0m}{2e}-\frac{1}{2}(\frac{1}{s}-\frac{1}{m}) $
and $n>>1$ (for details see \cite{gw}). From existing estimates of
the  bound on $\theta$ one can get an approximate value of the
binding energy. In a similar way, results for the scalar case can
also be obtained with $\nu =\frac{\theta_0m}{2e}-\frac{1}{2s}$.
\vskip .2cm \noindent
{\it Conclusion and future prospects}:\\
Let us conclude the paper with emphasizing the following point.
Effects of magnetic coupling qualitatively changes the behavior of
charged particles in the problem that we have considered, {\it
i.e.} scalar and spinor matter coupled to Maxwell-Chern-Simons
gauge fields. Especially, in case of scalars, NC effects are
solely responsible for the tantalizing possibility generating
Cooper pair like bound states. As was shown before \cite{gw},
these effects can be generated via the introduction of a {\it
non-minimal} gauge coupling. On the other hand, we have shown in
the present Letter that similar effects can be simulated if one
extends the interacting model (with {\it minimal} gauge coupling)
to the noncommutative plane. In the light of quantum gravity and
string theory results,  generalization of ordinary spacetime to a
noncomutative one is natural and physically motivated. It is also
interesting to note that a high energy effect such as
noncommutativity in spacetime can influence a low energy
phenomenon in a qualitative way. Hence it appears to us that
inducing the Pauli magnetic coupling effects by extending the
model to noncommutative space is a better option than directly
introducing a non-minimal gauge coupling at the fundamental level.

Lastly, we provide a  list of some of the interesting aspects of
the problem that
can be studied in near future:\\
1. Stern \cite{js} has shown that for a critical value of the
magnetic coupling (Pauli) term, the system reduces to that of free
anyons such that the electric effects (generated by the Maxwell
term) gets cancelled by the Pauli term contribution. Whether such
a thing occurs in our model and what is the subsequent critical
value of $\theta $ is an open problem. This issue is non-trivial
since the models in question, (that of \cite{js} and ours), are
not
identical.\\
2. Effects of space-time noncommutativity $(\vec{\theta} \ne 0)$
can be explicitly studied from the expression of our potential
(\ref{10},\ref{11}). Loss of unitarity in spacetime NC theories
and subsequent theoretical bounds  on partial wave amplitudes,
along the lines of \cite{chai}, can be studied without invoking
the non-relativistic limit. As expected, the space-time
noncommutativity destroys rotational invariance in the potential
(\ref{13}) whereas it remains intact with only spatial
noncommutativity. This is because in $2+1$-dimensions, $\theta _0$
points along the time direction, normal to the
plane \cite{sub}.\\
3. The same formalism can be applied in  $3+1$-dimensions where
the noncommutative effects
on the electrostatic potential can be studied.\\
4. Studies on Lorentz invariance violating interactions in a
different context has been
reported \cite{mmf}. Generic features of these models can be compared.\\
5. Extension of the model to higher orders in $\theta $- the
noncommutativity parameter and promoting to non-abelian gauge
groups can be rewarding. \vskip .2cm \noindent
{\it Acknowledgement}:\\
It is a pleasure to thank the Referee for the critical as well as
helpful comments.

\end{document}